\documentclass[osajnl2,showpacs,
twocolumn,%
superscriptaddress]{revtex4}  
\usepackage{hyperref}
\usepackage{textcomp}
\usepackage{amsmath}

\newcommand{\ket}[1]{{\lvert#1\rangle}}
\newcommand{\vect}[1]{{\boldsymbol{{\mathbf{#1}}}}}
\newcommand{\V}{\vect}
\newcommand{\prn}[1]{\left(#1\right)}

\begin{document}

\title{Modeling of pulsed laser guide stars for the {Thirty Meter Telescope} project}

\author{S.~M.~Rochester}
\email{simon@rochesterscientific.com} \affiliation{Rochester
Scientific, LLC, 2041 Tapscott Ave., El Cerrito, CA 94530, USA}
\author{A. Otarola}
\affiliation{TMT Observatory Corporation, 1200 East California
Boulevard, Mail Code 102-8, Pasadena, CA 91125, USA}
\author{C. Boyer}
\affiliation{TMT Observatory Corporation, 1200 East California
Boulevard, Mail Code 102-8, Pasadena, CA 91125, USA}
\author{D. Budker}
\affiliation{Department of Physics, University of California,
Berkeley, 366 LeConte Hall MC 7300 Berkeley, CA 94720-7300,
USA}%
\affiliation{Rochester Scientific, LLC, 2041 Tapscott Ave., El
Cerrito, CA 94530, USA}
\author{B. Ellerbroek}
\affiliation{TMT Observatory Corporation, 1200 East California
Boulevard, Mail Code 102-8, Pasadena, CA 91125, USA}
\author{R.\ Holzl\"{o}hner} \affiliation{European Southern Observatory (ESO), Garching bei M\"{u}nchen,
D-85748, Germany }
\author{L. Wang}
\affiliation{TMT Observatory Corporation, 1200 East California
Boulevard, Mail Code 102-8, Pasadena, CA 91125, USA}

\begin{abstract}
The Thirty Meter Telescope (TMT) has been designed to include
an adaptive optics system and associated laser guide star (LGS)
facility to correct for the image distortion due to Earth's
atmospheric turbulence and achieve diffraction-limited imaging.
We have calculated the response of mesospheric sodium atoms to
a pulsed laser that has been proposed for use in the LGS
facility, including modeling of the atomic physics, the
light-atom interactions, and the effect of the geomagnetic
field and atomic collisions. This particular pulsed laser
format is shown to provide comparable photon return to a
continuous-wave (cw) laser of the same average power; both the
cw and pulsed lasers have the potential to satisfy the TMT
design requirements for photon return flux.
\end{abstract}


\maketitle

\section{Introduction and motivation}

The Giant Segmented Mirror Telescope (GSMT) initiative has been
ranked among the highest priorities for US astronomy in the
last two decadal surveys of astronomy and astrophysics
\cite{DecadalSurvey2001,DecadalSurvey2010}. This ranking is
motivated by the potential scientific impact of the sensitivity
and diffraction-limited resolution that it is possible to
achieve with a large-aperture telescope. The Thirty Meter
Telescope (TMT) project has designed and is planning the
construction of a diffraction-limited, large-aperture segmented
telescope on the northern slope of the Mauna Kea volcano in
Hawaii (characteristics of the site are summarized in 
\cite{Schock2009}). For such a telescope, adaptive optics (AO)
and an associated laser guide star (LGS) system are needed to
achieve diffraction-limited capacity.

Adaptive-optics systems are of increasing importance for
current
\cite{Gavel2003,LeLouarn2004,Wizinowich2006,Boccas2006,Hart2010}
and future \cite{LloydHart2006,Ellerbroek2008,Diolaiti2008}
ground-based astronomical telescopes. Large ground-based
telescopes require AO to correct for the image distortion
induced by atmospheric turbulence. The use of AO requires a
reference light source, such as a bright natural star, in order
to measure the effect of the atmosphere on the light wavefront.
Often there is no bright natural star available within the
isoplanatic angle from the astronomical target, i.e., the angle
out to which phase variations in the wavefront are less than
one radian. In this case, a laser can be used to generate an
artificial guide star. This can be done by exploiting
Rayleigh-backscattered light from a pulsed laser
\cite{Thompson1992}, or in the case considered here, resonance
fluorescence from sodium atoms in the mesosphere
\cite{Jeys1991,Happer1994} at 80\,km to 125\,km altitude,
excited by either continuous-wave (cw) or pulsed lasers
resonant with the sodium D$_2$ transition at 589\,nm. Such
techniques were first proposed for astronomy in the 1980s
\cite{Foy1985,Fugate1991}; almost 30 years later, laser guide
star systems are available in all 6--10-m-class telescopes.

Many of the key scientific questions to be addressed with TMT
\cite{Steidel2009}---the nature and composition of the
universe, the formation of first stars and proto-galaxies, the
evolution of galaxies and the interplanetary medium, the
relation between black holes and their host galaxies, the
formation of stars and planets, the nature of extra-solar
planets, and the detection of extra-solar planets within the
habitable zone of their host stars---require
diffraction-limited imaging and thus a LGS-AO system. An
example of the need for diffraction-limited imaging is
observations of the crowded star field close to our galactic
center. Such observations are technically challenging and
astronomers have devised observation strategies to resolve the
stellar dynamics and accurately locate and weigh the
super-massive black hole in the center of our galaxy. Accurate
determinations of the stellar dynamics have evolved
significantly since the first survey of this star field by
means of speckle imaging in the late 1990s \cite{Matthews1996},
followed by Natural Guide Star-AO in the early 2000s
\cite{Schodel2002}; the best results have been achieved with
the use of LGS-AO in the mid- to late 2000s \cite{Ghez2008}.
Yet, there remains the need to improve these observations to
resolve stars at angular distances closer to the galactic
center. This will allow testing the predictions of general
relativity under the gravitational field of the super-massive
black hole hosted in the center of our galaxy. This will be
made possible by a large aperture, LGS-AO-assisted telescope,
such as TMT.

The TMT LGS Facility (LGSF) will initially include six lasers
to form an asterism consisting of one LGS on-axis and five LGS
equally spaced on a circle of 35\,arcsec angular diameter, with
the laser-launch telescope located behind the secondary mirror
\cite{Gilles2010}. Using the Multi-Conjugate Adaptive Optics
(MCAO) approach, the LGSF will be able to provide
diffraction-limited imaging in a field of view from tens of
arcseconds to about 1--2\,arcmin \cite{Wang2010}. The LGSF
design supports expansion to up to nine lasers to form
asterisms for other AO modes and radii from 5\,arcsec up to
510\,arcsec \cite{Boyer2010}.

The original design for the TMT LGSF called for cw lasers
(either solid-state Nd:YAG sum-frequency lasers or Raman fiber
lasers). Subsequently, TMT has also included the option of
using pulsed lasers based on diode-pumped solid-state
sum-frequency Nd:YAG technology. Such lasers have been
developed by scientists at the Technical Institute of Physics
and Chemistry of the Chinese Academy of Sciences (TIPC). In
this article, we investigate the effectiveness of pulsed lasers
for LGS, focusing on the TIPC laser format. We also compare the
photon return from the pulsed laser to that of a cw laser of
the same average power.


The effectiveness of an LGS system in correcting the
turbulence-induced aberrations in the image wavefront depends
on having sufficient photon return from the laser beacons. The
main requirements set for the first-light TMT AO system are the
following: (1) a root-mean-squared (rms) wavefront error at
zenith smaller than 187\,nm on-axis and 191\,nm over a
17\,arcsec field-of-view, (2) 50\% sky coverage at the galactic
pole, (3) operation up to a zenith angle of 65$^\circ$, and (4)
operation under atmospheric turbulence conditions characterized
by an atmospheric length scale (Fried length $r_0$
\cite{Fried1966}) as low as 0.1\,m in the direction of
observation. The TMT wavefront reconstruction algorithm
\cite{Gilles2006} has been tested under various scenarios and
is expected to meet the maximum
wavefront error specifications (see for instance 
\cite{Gilles2010}), with a significant margin for error, when
using a LGS that provides 900 photo-detected electrons (PDE)
per wavefront sensor sub-aperture in an integration time of
1.25\,ms. The specification of 900\,PDE applies when the laser
is pointed in the zenith direction, and includes a roughly
factor of two safety margin to account for various effects that
have the potential to increase the rms wavefront error,
including the effect of directing the laser at larger zenith
angles. The photon return scales due to purely geometrical
considerations as the cosine of the zenith angle (i.e., the
inverse of the airmass); this scaling is used as a reference
for comparison with the actual calculated directional
dependence of photon return reported here. Some additional
factors that affect the actual PDE count are the laser format
(power, polarization, spectral characteristics, etc), various
efficiencies in the propagation of the laser light to the
launch telescope, atmospheric absorption, the actual column
density of sodium atoms in the mesosphere, and the relative
direction of the local geomagnetic field with respect to the
laser beam. Also important is the efficiency of the LGS
detection system (i.e., the wavefront sensor).

Optimizing a laser system to maximize the guide star photon
return has been an area of ongoing research for many years.
Reference \cite{Hol2010} investigated the physical aspects
affecting the performance (i.e., photon flux return) of cw
sodium lasers. The calculation included the important physical
effects of Larmor precession of the sodium atoms in the
geomagnetic field, radiation pressure (recoil), and saturation
of the optical transitions, as well as velocity-changing and
spin-randomizing collisions. This was achieved using a
density-matrix calculation with coupled velocity groups. A
discussion was given of various mechanisms that are detrimental
to photon return, i.e., Larmor precession reduces the
efficiency of optical pumping, recoil depopulates the velocity
classes that interact with the laser light, and optical
saturation leads to stimulated emission, causing photons to be
emitted along the propagation of the laser rather than back to
the telescope.

Pulsed-laser LGS systems, in the general case in which the
pulses are not very short or very long, are more challenging to
model, due to the dynamical processes occurring on different
time scales. Experiments and modeling efforts have been
reported Refs.\ \cite{Avicola1994,Milonni1998,Pique2003}, among
others. However, none of the theoretical studies to date have
included all of the effects discussed above for the cw case,
which are all relevant to pulse lengths on a 100\,\textmu s
time scale. In addition, a full characterization of a
particular pulsed laser depends on its pulse length and pulse
repetition rate and how these parameters compare to collision
and relaxation rates and evolution time scales in the system,
as well as the multi-mode and coherence characteristics of the
laser.

In this paper, we report the results of physical modeling of
LGS with pulsed lasers, including all of the above-mentioned
effects. We focus on the TIPC pulsed laser and compute the
expected photon return at the TMT site as a function of pulse
length, pulse repetition rate, angular distance between the
laser propagation and the direction of the geomagnetic field.
As the first field test of the pulsed laser is planned for
mid-2012 at the location of the Large Zenith Telescope (LZT)
operated by the University of British Columbia in Canada, we
also perform simulations under the conditions at the LZT site.
In addition, the photon return of the pulsed laser is compared
to that obtained with a cw laser of similar transmitted power.
Briefly, we find that both the cw and pulsed formats appear to
be able to meet the design requirements for photon return. The
cw and pulsed formats provide comparable photon return, with
the differences at the level of the uncertainty in the model.

Results obtained from the numerical model also allow us to
investigate various physical mechanisms occurring in the system
that affect the photon return, often by aiding or inhibiting
the generation of ground-state atomic polarization. These
mechanisms can lead to interaction between seemingly unrelated
parameters of the system; for example, we find that the optimal
laser line width depends on the angle between the light
propagation direction and the geomagnetic field direction,
apparently due to the effect that each of these parameters has
on the creation of atomic polarization.

The paper is organized as follows. In Sec.\
\ref{sec:physicalsystem} the system being modeled and various
physical mechanisms that can influence the photon return are
discussed. In Sec.\ \ref{sec:descriptionofmodel} the details of
the Bloch-equation model and the numerical solver are given.
Section \ref{sec:inputdata} presents the environmental data and
the geometrical optics results that are used as inputs to the
model. In Sec.\ \ref{sec:integration} the method of integration
over the spatial extent of the beam is described. Section
\ref{sec:results} gives the results, and conclusions are given
in Sec.\ \ref{sec:conclusions}.

\section{Description of the system}
\label{sec:physicalsystem}

When sodium atoms are subject to a strong circularly polarized
light field resonant with the D$_2$ transition, a number of
physical processes occur that affect the efficiency at which
light is spontaneously re-emitted back toward the light source.

\begin{figure*}[tbp]
\centering
\includegraphics[width=5in]{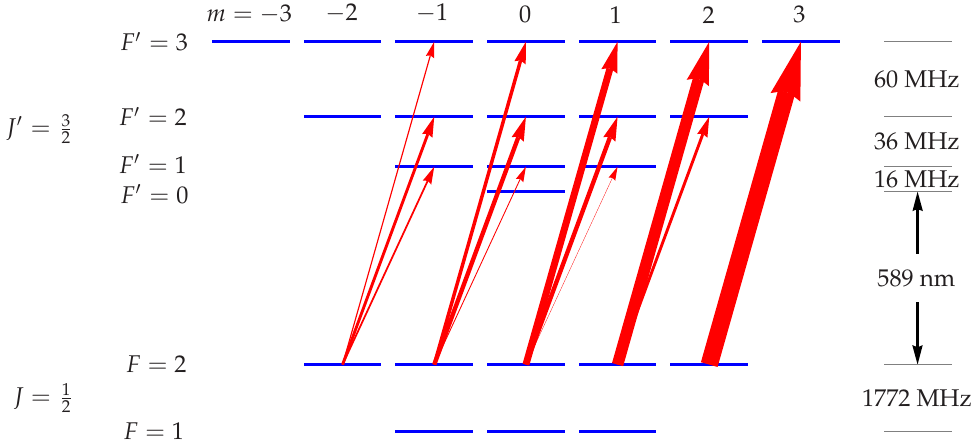}
\caption{(Color online) Level diagram of the Na D$_2$ line,
showing transitions induced by circularly polarized light
resonant with the D$_2$a transition. The widths of the arrows
indicate the relative transition strengths.
RochesterFig1.eps}\label{fig:D2LevelDiagramLight}
\end{figure*}

Left-circularly polarized light resonant with the D$_2$a
$F=2\rightarrow F'$ transition group induces optical pumping
that tends to transfer the atoms to the $\ket{F=2,m=2}$ ground
state (Fig.\ \ref{fig:D2LevelDiagramLight}). This atomic
ground-state polarization is beneficial for photon return in
three ways:
\begin{enumerate}
\addtolength{\itemsep}{-.5em}
\item The cycling $\ket{F=2,m=2}\rightarrow\ket{F'=3,m=3}$
    transition is the strongest transition in the D$_2$
    transition group, increasing the effective absorption
    cross section.
\item Atoms excited on the cycling transition cannot
    spontaneously decay to the $F=1$ ground state, where
    they would no longer interact with the light.
\item Fluorescence from $\ket{F'=3,m=3}$ is preferentially
    directed along the light beam, leading to an
    enhancement of photon flux observed at the location of
    the light source.
\end{enumerate}

Several mechanisms work to counteract the production of atomic
polarization---the relative importance of these is related to
the characteristic time scale of each.

\begin{figure}[tbp]
\centering
\includegraphics{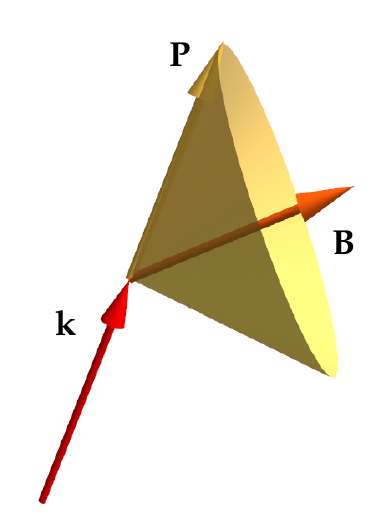}
\caption{(Color online) Precession of the atomic polarization $\V{P}$,
originally produced along the light wave vector $\V{k}$, about
the magnetic field vector $\V{B}$. If the polarization is
continuously generated and has a lifetime longer than the
Larmor precession time, it will tend to be averaged about the
magnetic-field direction. RochesterFig2.eps}\label{fig:LarmorPrecession}
\end{figure}

The geomagnetic field induces Larmor precession, which can act
to wash out the atomic polarization. Because circularly
polarized light produces atomic polarization parallel to the
light propagation direction, precession will be induced if the
magnetic field has a component transverse to the wave vector
$\V{k}$ (Fig.\ \ref{fig:LarmorPrecession}). A magnetic field of
0.33\,G, present at the Mauna Kea site, corresponds to a Larmor
precession frequency of $\sim\!220$\,kHz or a precession time
of 4.3\,\textmu s, meaning that this effect is important during
the 120\,\textmu s pulse produced by the TIPC laser. The
detrimental effect of the magnetic field is strongly dependent
on the direction of light propagation relative to the the
magnetic-field direction.

The atomic ground-state polarization is also relaxed by
spin-randomization collisions, predominantly with O$_2$
molecules \cite{Kibblewhite2009Maui}. The time scale for these
collisions varies with the O$_2$ density, but can be estimated
to be in the 200--5000\,\textmu s range \cite{Hol2010}. This
mechanism can thus weakly affect the dynamics during a
120\,\textmu s pulse. Finally, transit of the Na atoms through
the light beam (due to the motion of the atoms or the motion of
the beam itself) can also be considered to be a relaxation
mechanism. The net effective atomic velocity can be estimated
to be less than 370\,m/s \cite{Hol2010}, which, for a typical
beam size of 20\,cm, corresponds to a millisecond time scale
for transit relaxation. This effect can thus generally be
neglected for the pulse lengths in question, although we
include it in the numerical model for completeness.

The effective loss of atoms by spontaneous decay to the $F=1$
ground state can be directly combatted by applying ``repump''
light resonant with the D$_2$b $F=1\rightarrow F'$ transition
group to transfer the atoms back to the $F=2$ ground state
\cite{Jeys1991,Kibblewhite2009Maui,Hol2010}. As the simulation
results show, this technique is vital to obtaining sufficient
photon return fluxes. Alleviation of the loss of atoms to the
$F=1$ state also has the indirect effect of reducing the effect
of the magnetic field on the return flux, as this is one of the
worst effects resulting from the destruction of atomic
polarization caused by Larmor precession.

Another process that tends to reduce the photon return flux for
strong light fields is recoil of the atoms induced by
interaction with the light. Each absorption/spontaneous
emission cycle gives the atom a small momentum kick in the
laser beam propagation direction; the cumulative effect of
these kicks is to create holes in the Doppler distribution,
depleting the atomic velocity groups that are the most strongly
resonant with the light (this is illustrated in Sec.\
\ref{sec:velocitydependence}). Velocity-changing collisions of
the Na atoms with N$_2$ and O$_2$ molecules act to
reequilibrate the atomic velocity distribution, filling in the
holes created by recoil. These collisions occur on a time scale
of 15--300\,\textmu s \cite{Hol2010}, depending on gas density,
and so can act during a single laser pulse.

\section{Description of the model}
\label{sec:descriptionofmodel}

\subsection{The Bloch equations}

In order to calculate the observed fluorescence, the evolution
of the atoms is modeled using the optical Bloch equations for
the atomic density matrix. The density matrix describes the
magnitudes of and correlations between basis wave functions in
the ensemble of Na atoms, in particular, the spin wave
functions of the states involved in the Na D$_2$ transition. In
order to account for atoms with different Doppler shifts, the
density matrix is considered to be a function of atomic
velocity along the laser beam propagation direction. (An
additional degree of freedom may be included to account for the
multimode structure of the laser line as discussed below.) The
calculation is semiclassical in the sense that while the atoms
are treated quantum mechanically, the light fields are treated
classically; thus, the effect of spontaneous decay must be
included phenomenologically. Because the density matrix
describes all populations of, and coherences between, the 24
Zeeman sublevels making up the ground and excited states, the
calculation describes, in principle, all saturation and mixing
effects for essentially arbitrarily large optical and magnetic
fields. (In practice, certain coherences in the system are
negligible under normal experimental conditions and can be
neglected in order to increase the computational efficiency.)

In order to perform numerical calculations, the velocity
dependence of the density matrix is discretized to describe an
appropriate number $n_{v.g.}$ of velocity groups, each with a
fixed longitudinal velocity. Because coherences between atoms
with different velocities can be neglected, the complete
density matrix $\rho$ can be thought of as a collection of
$n_{v.g.}$ separate but coupled density matrices, each of
dimension $24\times24$.

The evolution of the density matrix is given by a
generalization of the Schr\"odinger equation:
\begin{equation}\label{Leq}
    \frac{d}{dt}\rho = \frac{1}{i\hbar}[H,\rho] + \Lambda(\rho) + \beta.
\end{equation}
Here the atomic level structure and the interaction with
external fields are described by the total Hamiltonian
$H=H_0+H_E+H_B$, with $H_0$ the Hamiltonian for the unperturbed
energy structure of the atom, $H_E=-\V{d}\cdot\V{E}$ the
Hamiltonian for the interaction of the electric dipole $\V{d}$
of the atom with the electric field $\V{E}$ of the light,
$H_B=-\V{\mu}\cdot\V{B}$ the Hamiltonian for the interaction of
the magnetic moment $\V{\mu}$ of the atom with the local
magnetic field~$\V{B}$, and the square brackets denote the
commutator. The term $\Lambda$ in Eq.\,(\ref{Leq}) represents
phenomenological terms added to account for relaxation
processes not described by the Hamiltonian. In our case these
relaxation processes include spontaneous decay (omitted from
the Hamiltonian due to the semiclassical approximation),
collisional spin relaxation (``S-damping'') proportional to
$S^2\rho-S\cdot(\rho S)$ \cite{Hap2010book}, where $S$ is the
electronic spin-angular-momentum operator, and the exit of
atoms from the light beam due to motion of the atoms and the
beam (transit relaxation). In addition, there are terms
included in $\Lambda$ to describe changes in atomic velocity
due to collisions and light-induced recoil, as well as an
effective relaxation rate for optical coherences that simulates
a laser spectrum with non-negligible bandwidth. These terms are
described in more detail below. {Each relaxation process
described by $\Lambda$ includes a corresponding
``repopulation'' process, so that the trace over the density
matrix for all velocity groups is conserved, corresponding to
conservation of the total number of atoms. The repopulation
process describing the entrance of atoms into the beam is
independent of $\rho$ and so is written as a separate term
$\beta$.}

Velocity-changing collisions are treated as hard collisions in
which the velocity of the colliding atom is rethermalized in a
Maxwellian distribution (no speed memory). The internal state
of the atom is assumed to be unchanged.

Light-induced recoil is described phenomenologically by causing
a fraction $v_r/\Delta v_{v.g.}$ of the excited-state atoms in
each velocity group to be transferred upon decay into the next
higher velocity group. Here $v_r$ is the recoil velocity and
$\Delta v_{v.g.}$ is the width of the particular velocity
group. This model relies on the fact that $v_r=2.9461$\,cm/s
(equivalent to a Doppler shift of $50$\,kHz) is much smaller
than the typical value of $\Delta v_{v.g.}$.

Equation~(\ref{Leq}) supplies a linear system of differential
equations for the density matrix elements, known as the optical
Bloch equations. Thinking of $\rho$ as a column vector of
$n_{v.g}\times24^2$ density matrix elements, the Bloch
equations can be written as $\dot\rho=A\rho+b$, where $A$ and
$b$ are a matrix and a vector, respectively, that are
independent of~$\rho$. The vector $b$ corresponds to $\beta$
and $A$ to the rest of the right-hand side of Eq.\ (\ref{Leq}).

The laser light field has a frequency component tuned near the
D$_{2}$a transition group, and may have an additional `repump'
component tuned near the D$_{2}$b transition group. Thus the
matrix $A$ has components that oscillate at each of these
frequencies. Under the rotating-wave approximation, the overall
optical frequency is removed from~$A$. However, the beat
frequency between the two light-field components remains. This
beat frequency can also be removed from the Bloch equations in
our case: each frequency component interacts strongly with one
transition group and very weakly with the other, so the weak
coupling can be neglected for each transition. If, in addition,
the small magnetic-field-induced mixing between the two
hyperfine ground states is neglected, the beat frequency can be
entirely removed from the evolution equations.

The fluorescent photon flux per solid angle emitted in a given
direction can be found from the solution for $\rho$ as the
expectation value of a fluorescence operator \cite{Corney}.

\subsection{Methods for simulation of multiple laser modes}

The spectrum of the TIPC laser is composed of three equally
spaced modes. The proper method for including the effect of the
modes in the model depends on the details of the time
dependence of the spectrum, including the coherence time
between the modes and the pulse-to-pulse fluctuations of the
mode amplitudes. Because these details were not known with
certainty, three different methods were compared in the study.

In the ``coherent-modes'' method, the three modes are assumed
to be perfectly coherent and non-fluctuating. The atoms are
subject to a field with components oscillating at each of the
three mode frequencies. This is equivalent to a single field,
with a frequency equal to that of the central mode, amplitude
modulated with a modulation frequency equal to the mode
separation. Using this model, the resulting return-flux signal
exhibits beats at the modulation frequency. The width of each
laser mode can be accounted for by including additional
relaxation terms for the atomic optical coherences. We note
that even when the effective line widths are broadened so that
they overlap, the observed beats remain, because the modes are
still perfectly coherent. Direct integration of the evolution
equations in this situation is computationally slow, because of
the rapid oscillations in the solution. Therefore, we have
implemented a Floquet technique in which the explicit
appearance of the rapid oscillation is removed from the
evolution equations through expansion of the density matrix in
a Fourier series. This results in a coupled system of equations
for the relatively slowly evolving Fourier coefficients. These
equations can be efficiently integrated to find the time
dependence of the lowest few Fourier harmonics.

In the ``mode-hopping'' method, the laser frequency is assumed
to fluctuate between the three mode frequencies on a time scale
much shorter than the pulse length. This is accomplished by
writing density matrices describing ``regions,'' each subject
to one of the laser modes, and allowing atomic transit between
the regions. The transit rate corresponds to the width of each
laser mode. In this model, the optical coherences are assumed
to be destroyed upon transit between regions, meaning that the
laser modes can be considered to be effectively incoherent.
Thus no beats are observed in the return flux.

In the ``pulse-averaging'' method, large pulse-to-pulse
fluctuations in the relative mode amplitudes are assumed, so
that during each pulse the laser power is present on only one
of the three modes, with equal probability. This is
accomplished by running the simulation three times, once with
the light frequency equal to each mode frequency, and averaging
the results. The modes are effectively incoherent with this
method.


\subsection{The numerical solver}

The density-matrix evolution equations are generated using the
LGSBloch package for Mathematica, which is based on the Atomic
Density Matrix package written by S. Rochester \cite{ADMweb}.
The system of ordinary differential equations is solved using a
code written in C and based on the open-source ODE solver
CVODE, from the SUNDIALS package \cite{sundialsweb}. This
solver addresses the stiffness of the system using the implicit
backward differentiation formulas (BDFs) \cite{Hindmarsh2005}.
These methods require the solution of a linear system of
equations at each step of the solver; these solutions are
accomplished with a preconditioned Krylov method. (In such a
method an initial guess is improved by minimizing the residual
over a subspace with dimension much smaller than that of the
full system \cite{Hindmarsh2005}.) The rate of convergence of
the method is increased by pre-multiplication with a
block-diagonal preconditioner (approximate inverse of $A$),
obtained by setting all terms that connect density matrix
elements from different velocity groups to zero, and then
inverting the block for each velocity group. For the
mode-hopping multimode method, the transit terms that couple
the density matrices for the three modes are also included in
the preconditioner, using an operator splitting technique that
inverts these terms separately. For the coherent-modes Floquet
technique, the terms for all of the included coupled Fourier
coefficients are included in the block for each velocity group.

\section{Input data}
\label{sec:inputdata}

\subsection{Standard parameters and environmental conditions}

\begin{figure}[tbp]
\centering
\includegraphics{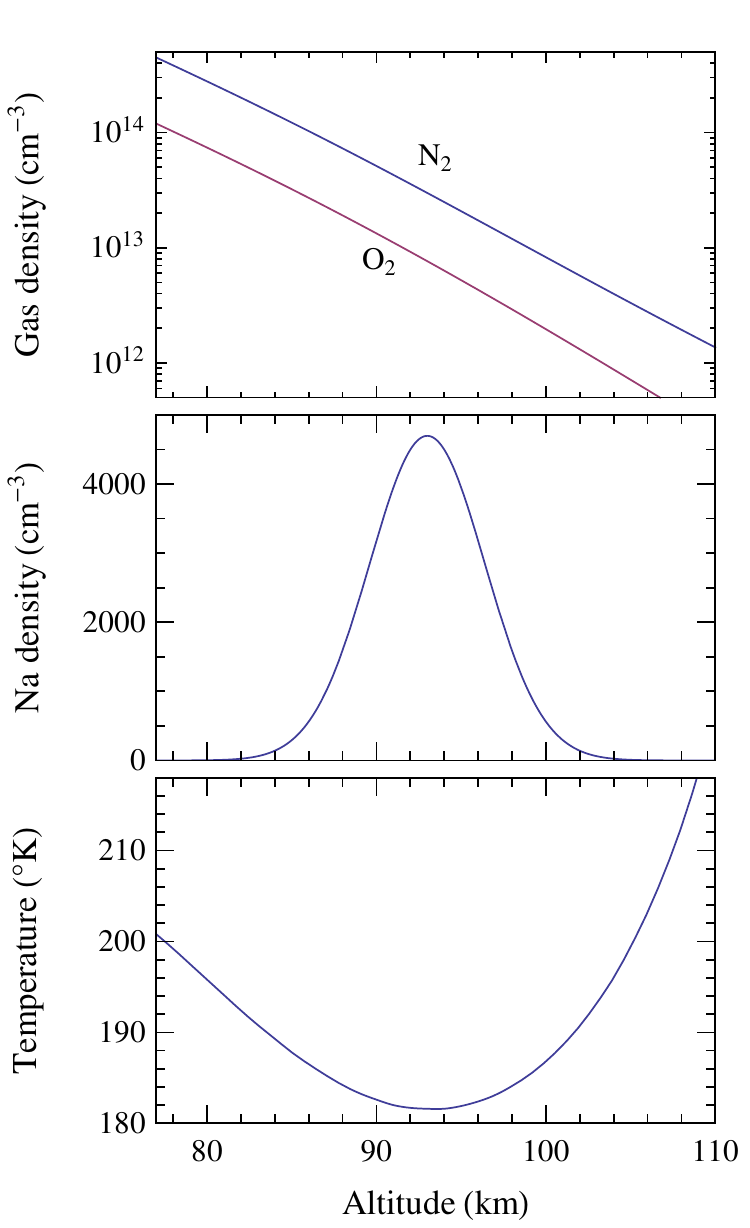}
\caption{(Color online) Temperature and O$_2$ and N$_2$ gas densities as a
function of altitude obtained from the MSIS-E-90 Atmosphere
Model for the Mauna Kea site, along with the assumed Na density
profile. (Note that deviations of the actual Na density profile
from a Gaussian are often quite
large.) RochesterFig3.eps}\label{fig:EnvironmentPlots}
\end{figure}

The standard parameters used in the model for the TIPC laser
are given in Table \ref{tab:laserparams}. Environmental
parameters and parameters specific to the Mauna Kea and LZT
sites are listed in Table \ref{tab:siteparams}. The sodium
density as a function of altitude is assumed to be described by
a Gaussian function centered at 93\,km with a full width at
half maximum (FWHM) of 8\,km. This is a rough estimate, as the
measured density profile at any given time may deviate
significantly from a Gaussian. However, our modeling indicates
that the photon return has only a weak dependence on the width
of the density profile: increasing the assumed width to 11\,km
leads to a reduction in photon return of about 1\%. For each
site, the temperature and gas densities as a function of
altitude in the sodium layer were obtained from the MSIS-E-90
Atmosphere Model \cite{MSISE90}; for the Mauna Kea site, these
values, along with the assumed Na density profile, are plotted
in Fig.\ \ref{fig:EnvironmentPlots}. Collision rates and the
Doppler width as a function of altitude are derived from these
data and the assumed collision cross sections given in Table
\ref{tab:crosssections}.

\begin{table}
\caption{Standard parameters for the TIPC pulsed
laser.}\label{tab:laserparams}
\begin{center}
\begin{tabular}{@{\extracolsep{10pt}}ll}\hline
$\text{Average laser power}$&$20\text{$\, $W}$\\
$\text{Average projected laser power}$&\\
\qquad Mauna Kea&$12\text{$\, $W}$\\
\qquad LZT&$18.6\text{$\, $W}$\\
Polarization & circular\\
$\text{Number of modes}$&$3$\\
$\text{Mode spacing}$&$150\, \text{MHz}$\\
$\text{Pulse length}$&$120\, \text{\textmu s}$\\
$\text{Repetition rate}$&$800\, \text{Hz}$\\
\hline
\end{tabular}
\end{center}
\end{table}

\begin{table}
\caption{Standard parameters for Mauna Kea and LZT sites. The
Fried length is the size of the circular aperture within which
the variance of the wavefront phase aberrations is less than 1
radian \cite{Fried1966}.}\label{tab:siteparams}
\setlength{\columnsep}{20pt}
\begin{center}
\begin{tabular}{@{\extracolsep{5pt}}lll}\hline
& Mauna Kea & LZT \\
\hline
Latitude & 19.8$^\circ$ & 49.3$^\circ$\\
Longitude & 155.5$^\circ$ & 122.6$^\circ$\\
$\text{Observatory altitude}$&$4050\text{$\, $m}$&$395\text{$\, $m}$\\
$\text{B-field}$&$0.334\text{$\, $G}$&$0.525\text{$\, $G}$\\
$\text{Na Larmor frequency}$&$234\, \text{kHz}$&$367\, \text{kHz}$\\
$\text{B-field zenith angle}$&$\text{126$ {}^{\circ}$18'}$&$\text{160${}^{\circ}$30'}$\\
$\text{B-field azimuthal angle}$&$\text{9$ {}^{\circ}$45'}$&$\text{16${}^{\circ}$57'}$\\
$\text{Atmospheric transmission (zenith)}$&$0.84$&$0.8$\\
Fried length (zenith) & 21\,cm & 5\,cm\\
$\text{Na column density}$&\multicolumn{2}{c}{$4.\times 10^{13}\text{$\, $m}^{-2}$}\\
$\text{Na centroid altitude}$&\multicolumn{2}{c}{$93\, \text{km}$}\\
Na layer thickness & \multicolumn{2}{c}{$8\, \text{km}$}\\
\hline
\end{tabular}
\end{center}
\end{table}

\begin{table}
\caption{Estimated values for cross-sections (see 
\cite{Hol2010} and references therein; note that the factor of
$1/2$ described in 
\cite{Hol2010} is incorporated into
$\sigma _{\text{Na}-\text{O}_2}^S$).}\label{tab:crosssections}
\begin{center}
\begin{tabular}{@{\extracolsep{10pt}}ll}\hline
$\sigma _{\text{Na}-\text{N}_2}$&$7.15\times 10^{\text{-15}}\, \text{cm}^2$\\
$\sigma _{\text{Na}-\text{O}_2}$&$7\times 10^{\text{-15}}\, \text{cm}^2$\\
$\sigma _{\text{Na}-\text{O}_2}^S$&$2.5\times 10^{\text{-15}}\, \text{cm}^2$\\[2pt]
\hline
\end{tabular}
\end{center}
\end{table}

\subsection{Laser spot image model}

To find the expected laser intensity profile, a geometrical
optics simulation was performed. The laser launch telescope
(LLT) has a diameter of 0.4\,m and is nominally focused (absent
diffraction and turbulence effects) at a range of 125\,km. The
Gaussian laser beam has a waist diameter ($e^{-2}$ of the peak
intensity) of 0.24\,m at the LLT pupil.


We computed the expected laser intensity profile at ranges of
85, 95, 105, 115, and 125\,km. This was accomplished by
modeling the propagation of the 589\,nm laser light from the
launch telescope for several instantiations of a single
turbulence screen assumed located close to the LLT. The
turbulence screen was modeled with a von-Karman power spectral
density, an outer scale of turbulence \cite{Tatarski1961} set
to 30\,m, and for various turbulence strengths characterized by
the Fried lengths $r_0=0.10$, 0.15, 0.20, 0.25, and 0.30\,m.
The TIPC geometrical beam quality is characterized by a
broadening factor of $M^2=1.2$. This $M^2$ factor is achieved
in the light propagation model by introducing a slight coma
aberration in the LLT mirror and a total wavefront error of
76.5\,nm rms. In addition, because the LLT mirror has a fixed
focal distance of $f_L=125$\,km, the laser intensity beam
profiles at altitudes lower than 125\,km also include the
effect of a focus error modeled as $r^2 \Delta h/(2f_L^2)$,
where $r$ is the laser beam radius and $\Delta h$ is the
distance from the LGS spot to the 125\,km focal distance.

The beam profiles generated through the process above were
interpolated as needed for intermediate values of $r_0$ used
for simulations at the Mauna Kea site. At the LZT site, the
atmospheric turbulence is stronger, with the median turbulence
conditions better characterized by an $r_0$ of 0.05\,m. In this
case the laser beam profile was extrapolated linearly down to
$r_0=0.05$\,m from the profiles available for larger values of
$r_0$.

\begin{figure}[tbp]\centering
\includegraphics{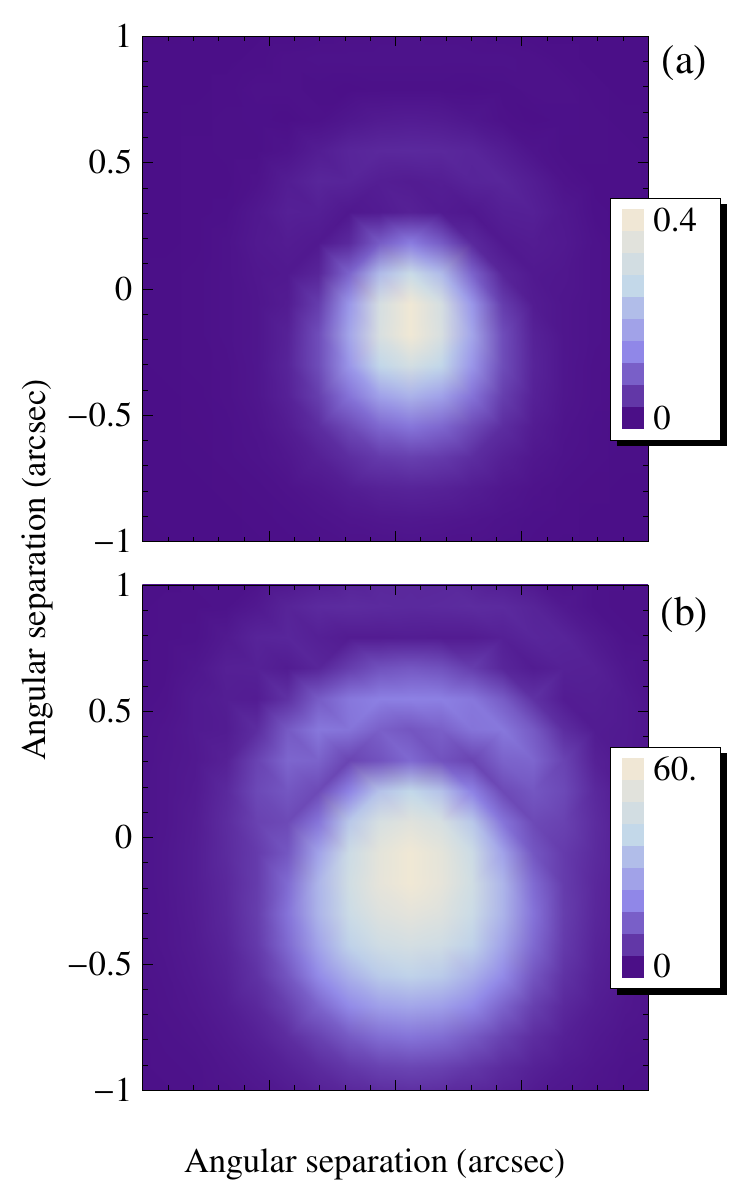}
\caption{(Color online) (a) Interpolated beam profile from geometrical optics
model, given in terms of the fraction of power in each pixel.
The peak value is the Strehl ratio. (b) Return flux from each
pixel in $10^3$ photons/s/m$^2$. RochesterFig4.eps}
\label{fig:BeamAndFluxProfile}
\end{figure}

The LGS spot images are sampled at 0.121" and are 64 pixels
across. An example of the beam profile is given in Fig.\
\ref{fig:BeamAndFluxProfile}(a). Fitting the narrower dimension
of the beam profiles with a Gaussian function yields angular
sizes ranging from 0.68" to 0.37" FWHM for $r_0=0.05$\,m to
$0.3$\,m.

\section{Spatial integration}
\label{sec:integration}

Integration of the observed return flux over the spatial extent
(profile and path length) of the laser beam is accomplished in
four steps, using a method similar to that of 
\cite{Holzlohner2010SPIE}:
\begin{enumerate}
\addtolength{\itemsep}{-.5em}
\item Construct table of return flux vs.\ intensity and
    altitude.
\item For each pixel in the modeled beam profile, propagate
    light through sodium layer using the table, keeping
    track of absorption.
\item Integrate return flux over path length.
\item Sum over pixels.
\end{enumerate}

We first generate a table of return fluxes as a function of
light intensity and altitude. Parameters that change with
altitude, plotted in Fig.\ \ref{fig:EnvironmentPlots}, are
N$_2$ and O$_2$ density (affects collision rates), Na density
(multiplicative factor), and temperature (weakly affects
collision rates and Doppler width). Sodium density is modeled
by a Gaussian with integrated column density
$4\times10^{13}\,\text{m}^{-2}$. N$_2$ and O$_2$ density and
temperature are obtained from the MSIS-E-90 Atmosphere Model
\cite{MSISE90}.

Both the return flux $\Phi(I,z)$ and total fluorescence into
4$\pi$, $\Phi_\text{tot}(I,z)$, are obtained from the model.
Return values from the model are multiplied by the Na density
and area of a 0.12" spot (the size of a pixel in the beam
profile) at a given altitude. This results in values of fluxes
per unit path length $d\ell$.

Light power in the uplink beam is absorbed as it travels
through the sodium layer, and reradiated as fluorescence in all
directions. For a sub-beam with angular size equal to one pixel
in the beam profile, the power absorbed per unit path length
$d\ell$ is $\hbar\omega d\Phi_\text{tot}(I,z)/d\ell$, with
$I=P(\ell)/(\theta_p\ell)^2$ and $z=\ell/X$, where $P(\ell)$ is
the power at distance $\ell$, $\theta_p$ is the angular size of
one pixel, $X=\sec\zeta$ is the airmass ($\zeta$ is the zenith
angle) \cite{curvaturenote}, and $\hbar \omega$ is the energy
of one photon. Thus the power along the optical path is given
by the propagation equation
\begin{equation}\label{eq:abseq}
    dP(\ell)/d\ell =
    -\hbar\omega d\Phi_\text{tot}(P(\ell)/(\theta_p\ell)^2,\ell/X)/d\ell.
\end{equation}
We integrate this equation for a range of initial light powers,
chosen to cover the range obtained using the given beam profile
and average laser power.

For the transmission of the return flux through the sodium
layer, we assume linear absorption and so use the curve
obtained from Eq.\ (\ref{eq:abseq}) in the low power limit,
denoted by $T_\text{lin}(\ell)$. The observed return flux into
a detector of unit area per unit path length emitted from a
pixel-sized sub-beam at a distance $\ell$ is then given by
\begin{equation}
    d\Phi_\text{obs}(P_0,\ell)/d\ell
    = \frac{T_a^X}{\ell^2}T_\text{lin}(\ell)
        d\Phi(P(\ell)/(\theta_p\ell)^2,\ell/X)/d\ell,
\end{equation}
where $P(\ell)$ depends on the initial power in the sub-beam,
$P_0$, and $T_a$ is the atmospheric transmission at zenith.

Integrating $d\Phi_\text{obs}(P_0,\ell)/d\ell$ over the path
length, we find the observed flux from each sub-beam as a
function of the initial power, $\Phi_\text{obs}(P_0)$. We then
sum the return flux from all of the pixels in the beam profile.
An example of the return flux profile is shown in Fig.\
\ref{fig:BeamAndFluxProfile}(b), the intermediate steps in the
integration procedure are illustrated in Sec.\
\ref{sec:integrationillustration}.

\section{Results}
\label{sec:results}

In Secs.\
\ref{sec:timedependence}--\ref{sec:velocitydependence}, we
illustrate various aspects of the signal using the return flux
calculated from a particular location in the sodium layer,
i.e., not integrated over the spatial extent of the beam. Sec.\
\ref{sec:integrationillustration} illustrates the steps taken
in the spatial integration procedure, Sec.\
\ref{sec:integrationresults} presents the integrated return
flux results, and Sec.\ \ref{sec:optimization} presents plots
that relate to optimization of the pulse length, repump power
fraction, and laser line width.

\subsection{Time dependence}
\label{sec:timedependence}

\begin{figure}[tbp]
\centering
\includegraphics{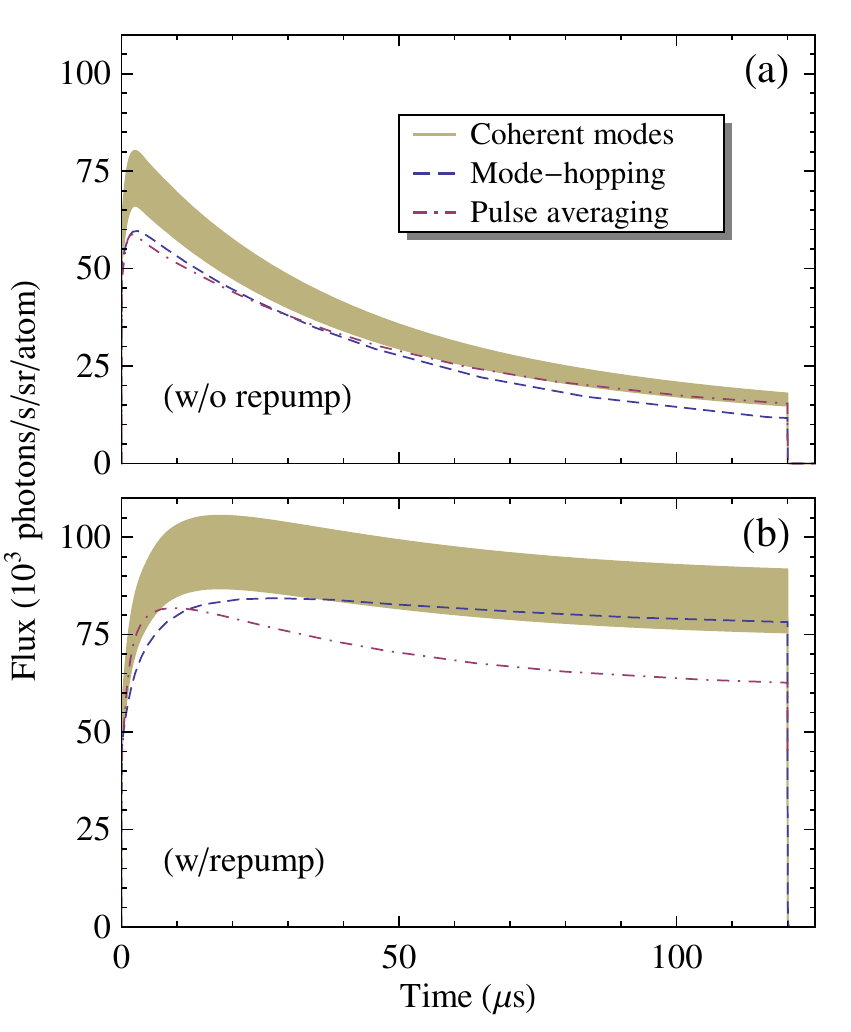}
\caption{(Color online) Instantaneous return flux during a
single pulse as a function of elapsed time, using three
different multimode treatments, both with no repump light and
with 10\% repump. Example plots are for Mauna Kea, zenith angle
$=30^\circ$, azimuth angle $=190^\circ$,
$I_\text{avg}=27.5$\,W/m$^2$. The oscillations present in the
coherent-mode treatment are too rapid to be distinguished on
this time scale, and so appear as solid bands.
RochesterFig5.eps}\label{fig:TimeDependencePlot}
\end{figure}

Examples of the time dependence of photon flux return during a
120\,\textmu s square pulse using the three different methods
for treating the multimode laser spectrum (coherent modes,
mode-hopping, and pulse averaging) are shown in Fig.\
\ref{fig:TimeDependencePlot}. In Fig.\
\ref{fig:TimeDependencePlot}(a), no repump light is applied.
Following the initial rapid rise in flux at the beginning of
the pulse, the flux begins to fall off. This is primarily due
to pumping to the $F=1$ ground state, as can be seen by
diverting some of the laser power into the repump light beam
[Fig.\ \ref{fig:TimeDependencePlot}(b)], which counteracts most
of the fall off. As throughout, unless otherwise noted, we take
10\% of the total laser power to be repump light. We also
assume that the repump light is generated by amplitude
modulation of the D$_2$a laser field, meaning that an
additional 10\% of the laser power is lost into the additional
low-frequency sideband that is produced. Unless otherwise
noted, the repump light is taken to be of the same polarization
as the main light field. The effect of the repump light is
examined in more detail in Sec.\ \ref{sec:velocitydependence}.

\begin{figure}[tbp]
\centering
\includegraphics{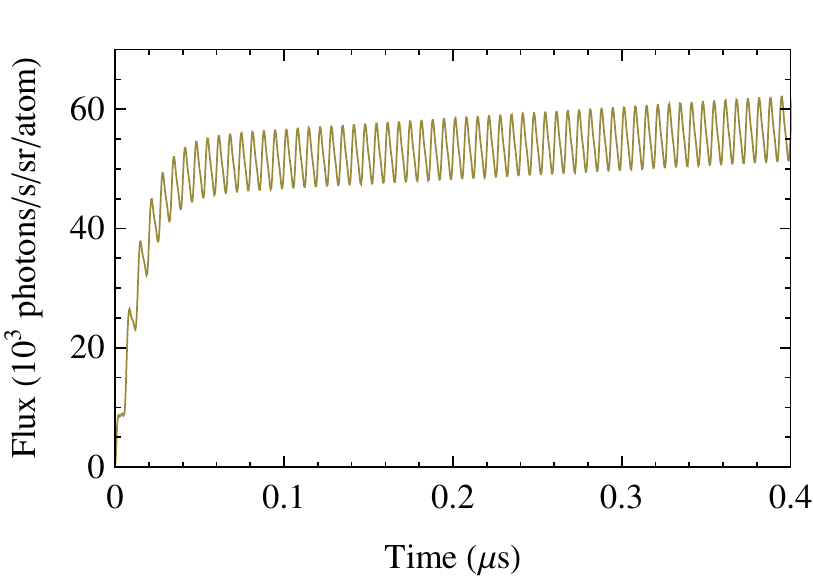}
\caption{(Color online) An expanded view of the initial part of the
coherent-modes plot from Fig.\
\ref{fig:TimeDependencePlot}(b). The oscillations are due to
interference between the laser
modes. RochesterFig6.eps}\label{fig:InitialTimeDependencePlot}
\end{figure}

The rapid beats present when using the coherent modes treatment
can be seen more clearly in Fig.\
\ref{fig:InitialTimeDependencePlot}, which shows the initial
part of the pulse.

\subsection{Pulse shape}

\begin{figure}[tbp]
\centering
\includegraphics{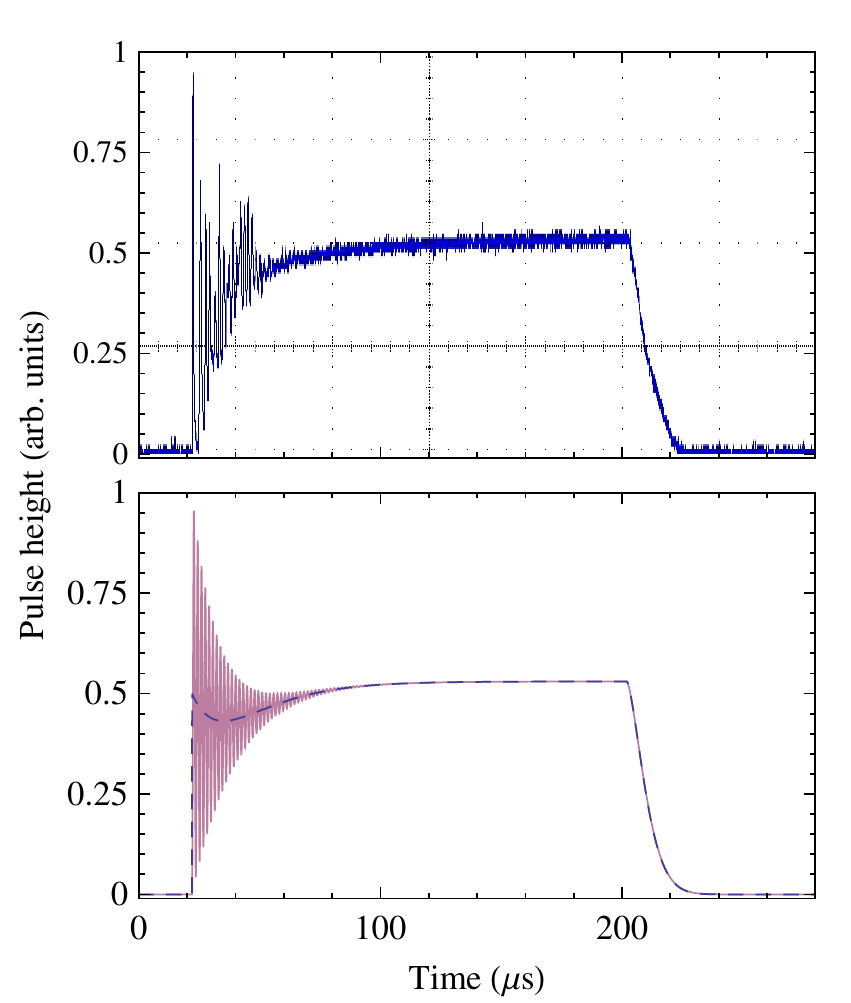}
\caption{(Color online) Laser pulse shapes: (top) The supplied
measured pulse shape; (bottom) the pulse-shape model described
in the text (solid line) and the same model with the initial
oscillations averaged out (dashed line). The parameters used
here are $t_0=22$\,\textmu s, $\tau_r=19$\,\textmu s,
$a_\text{osc}=1.9$, $\tau_\text{osc}=12$\,\textmu s,
$\omega=2\pi\times0.6$\,MHz, $t_1=202$\,\textmu s,
$\tau_f=9.6$\,\textmu s, and $p=1.5$. The overall amplitude is
scaled to match the measurement. The nominal pulse length is
$t_1-t_0=180$\,\textmu s, longer than the standard pulse length
of 120\,\textmu s used in the simulations.
RochesterFig7.eps}\label{fig:PulseShapeData}
\end{figure}

The results of Fig.\ \ref{fig:TimeDependencePlot} are
calculated assuming a perfectly square pulse shape. A
measurement of the actual pulse shape produced by the TIPC
laser is shown in Fig.\ \ref{fig:PulseShapeData}. This shape
differs from a square pulse in that there are finite rise and
fall times, and also short spikes in the initial part of the
pulse. To investigate the effect of these differences, we
defined an approximate analytical model for the measured pulse
shape. The model represents the initial spikes as rapid
exponentially decaying oscillations. While not a completely
faithful reproduction of the measured pulse shape, the accuracy
of this representation should be enough to indicate whether the
initial spikes have the potential to cause a significant
deviation from the results obtained using an ideal square
pulse.

To model the measured pulse shape, we used a functional form
with the pulse rise given by
\begin{equation}\label{eq:pulserise}
    1-e^{-(t-t_0)/\tau_r}+a_\text{osc}e^{-(t-t_0)/\tau_\text{osc}}\sin^2(\omega t/2),
\end{equation}
where $t_0$ is the pulse start time, $\tau_r$ is the rise time,
$a_\text{osc}$ is the amplitude of the initial oscillations,
$\tau_\text{osc}$ is the decay time of the oscillations, and
$\omega$ is the oscillation frequency. The pulse fall is
somewhat faster than exponential, and so is modeled by
\begin{equation}
    \exp\prn{-[(t-t_1)/\tau_f]^p},
\end{equation}
where $t_1$ is the time at the beginning of the fall, $\tau_f$
is the nominal fall time, and $p\approx1.5$.

Appropriate parameters were chosen by eye to match the supplied
measured plot, as shown in Fig.\ \ref{fig:PulseShapeData}. Also
shown in the figure is a pulse shape obtained by averaging out
the initial oscillations of the formula \eqref{eq:pulserise}.

\begin{figure}[tbp]
\centering
\includegraphics{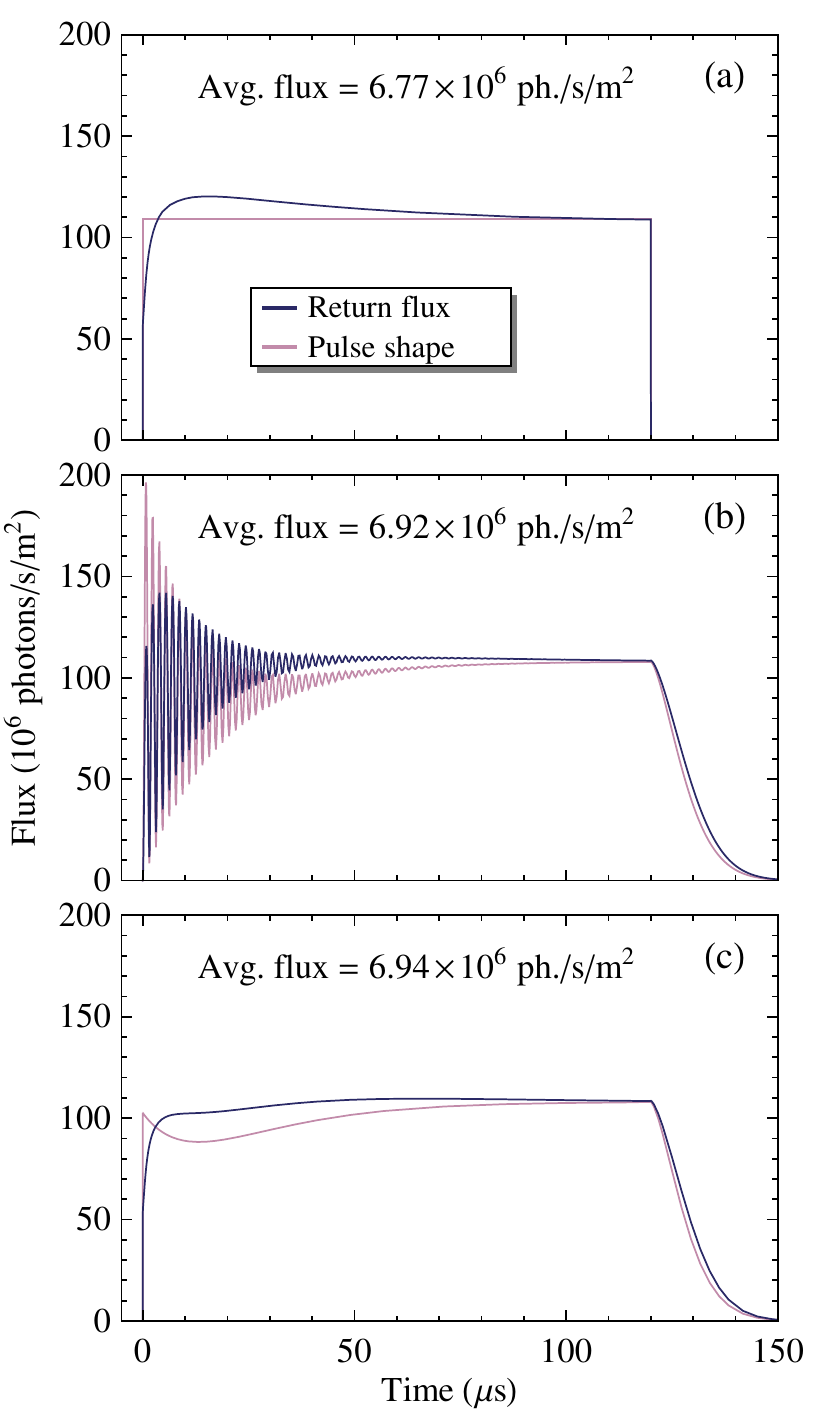}
\caption{(Color online) Return flux in response to three different pulse
shapes: (a) ideal square pulse, (b) pulse with initial
oscillations and finite rise and fall times, and (c) pulse as
in (b) but with initial oscillations averaged out. A
single-mode laser is assumed. The nominal pulse lengths are
120\,\textmu s; the other pulse-shape parameters are as given
in the caption to Fig.\
\ref{fig:PulseShapeData}. RochesterFig8.eps}\label{fig:PulseShape}
\end{figure}

The calculated responses to a square pulse, a pulse with
initial oscillations, and a pulse with the initial oscillations
averaged out are shown in Fig.\ \ref{fig:PulseShape}(a), (b),
and (c), respectively. The pulses are each 120\,\textmu s from
the beginning of the pulse to the start of the fall, with the
other pulse-shape parameters chosen to match those given in
Fig.\ \ref{fig:PulseShapeData}. The three types of pulses are
each normalized to have the same total pulse energy. A
single-mode laser is assumed for these plots.

The time-averaged flux from the more realistic pulse is higher
than that from the square pulse, apparently because the pulse
energy is spread out over a longer effective time due to the
pulse fall time. Averaging out the initial oscillations has
almost no effect on the time-averaged flux.

These results indicate that the deviations of the measured
pulse shape from a square pulse have an insignificant effect on
the photon return, given the current uncertainty in the
environmental and laser parameters. We therefore will use an
ideal square pulse in the following simulations.

\subsection{Atomic-velocity dependence}
\label{sec:velocitydependence}

\begin{figure}[tbp]
\centering
\includegraphics{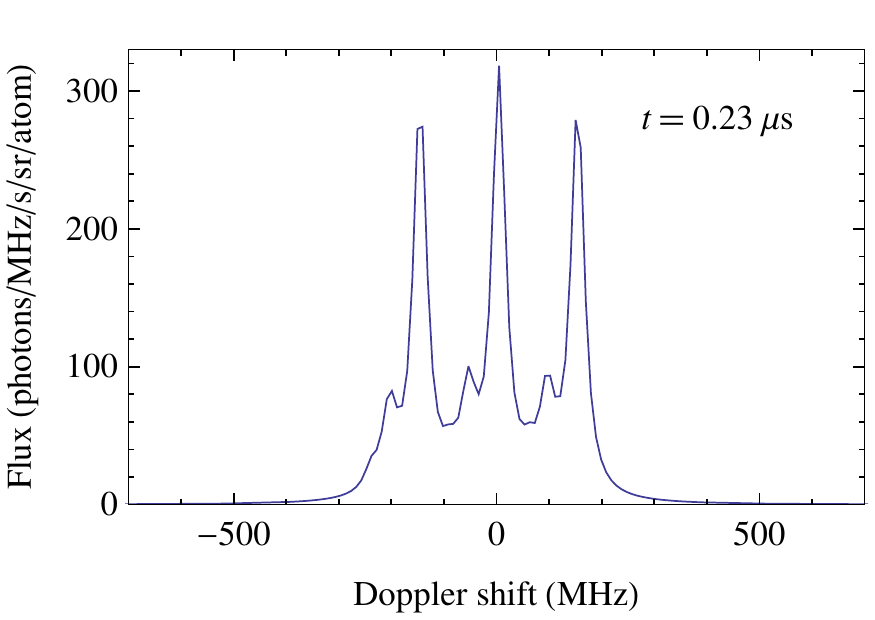}
\caption{(Color online) An example of return flux as a function of the longitudinal atomic
velocity (given in terms of the corresponding Doppler
shift), at a time near the beginning of a laser pulse. RochesterFig9.eps}\label{fig:VelocityDependence}
\end{figure}

The return flux as a function of the atomic velocity group near
the beginning of the laser pulse is shown in Fig.\
\ref{fig:VelocityDependence}. The response to the three laser
modes is seen, with the shape of each peak determined by the
upper-state hyperfine structure of the D$_2$ line.

\begin{figure*}[tbp]
\centering
\includegraphics{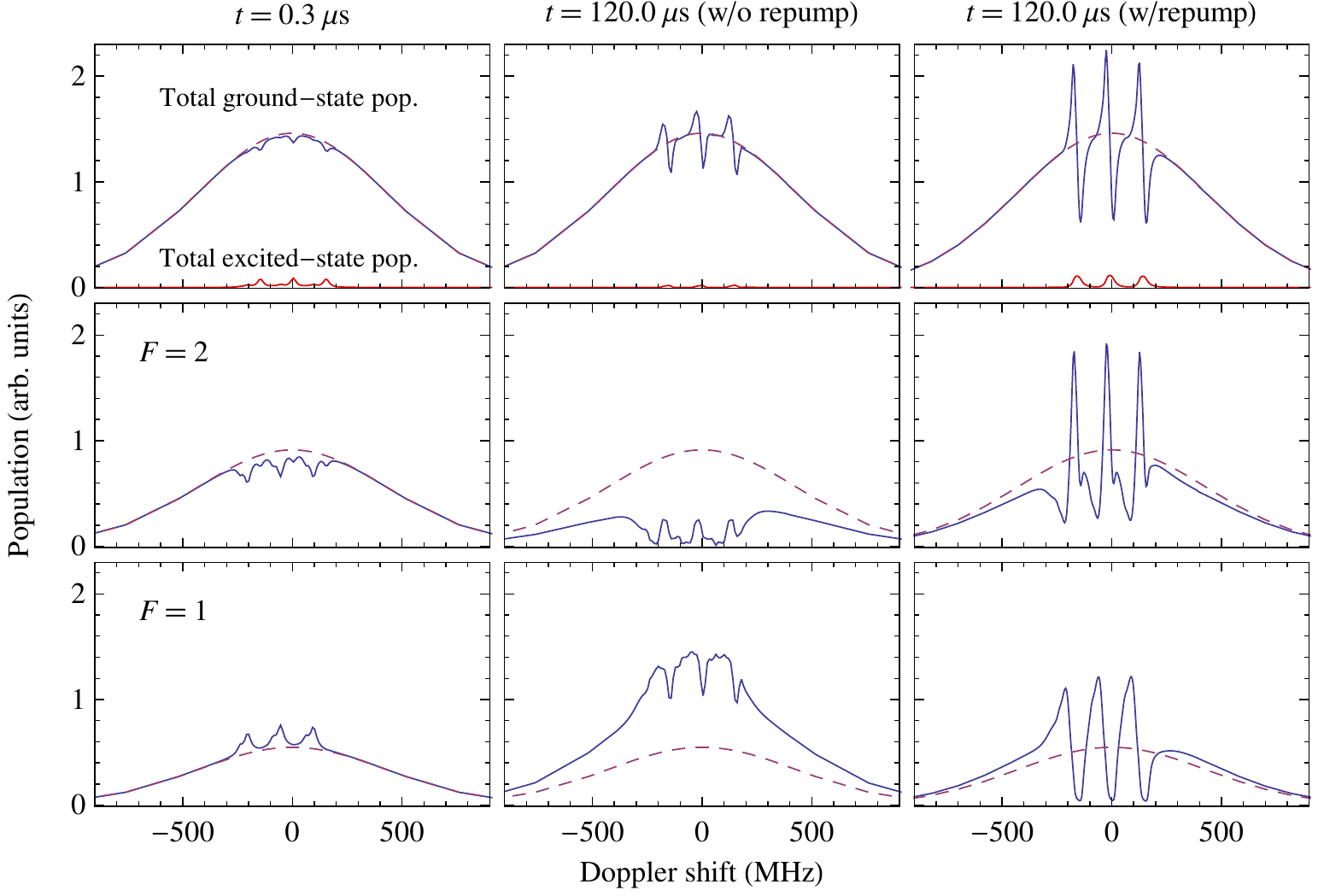}
\caption{(Color online) Populations of the $F=1$ ground state
(bottom row), $F=2$ ground state (middle row) and total
ground-state and upper-state populations (top row), as a
function of longitudinal atomic velocity for the mode-hopping
method. Populations are shown for times near the beginning of
the pulse (left column), and at the end of a pulse without
repump light (middle column), and at the end of a pulse with
repump light (right column), superimposed over the unperturbed
equilibrium populations (dashed lines).
RochesterFig10.eps}\label{fig:Populations}
\end{figure*}

To study the effect of various mechanisms on the dynamics of
the atomic system, it is helpful to plot atomic populations as
a function of velocity group at various times during the pulse.
Figure \ref{fig:Populations} shows populations of the $F=1$
ground state (bottom row), $F=2$ ground state (middle row) and
total ground-state and upper-state populations (top row), as a
function of longitudinal atomic velocity using the mode-hopping
treatment of the laser modes. The left-hand column shows the
populations near the beginning of the pulse. A small amount of
the total population has been transferred from the ground state
to the upper-state, and some atoms have been transferred from
the $F=2$ ground state to the $F=1$ ground state by optical
pumping. The middle column shows the populations at the end of
the pulse if no repump light is used. In this case the majority
of the $F=2$ atoms have been pumped into the $F=1$ ground
state, where they do not interact with the laser light. As a
result, the number of atoms in the excited state (and hence the
photon return flux) is greatly reduced. The hole-burning effect
of atomic recoil can also be seen in the total ground-state
population. The right-hand column shows what happens if repump
light is used. In this case, the population of the $F=2$ state
is largely preserved, as atoms pumped to the $F=1$ state are
repumped to the $F=2$ state. When repump light is applied the
effect of recoil becomes much more dramatic, as each atom is
able to interact with the light many more times during the
pulse.

\subsection{Spatial integration}
\label{sec:integrationillustration}

\begin{figure}[tbp]
\centering
\includegraphics{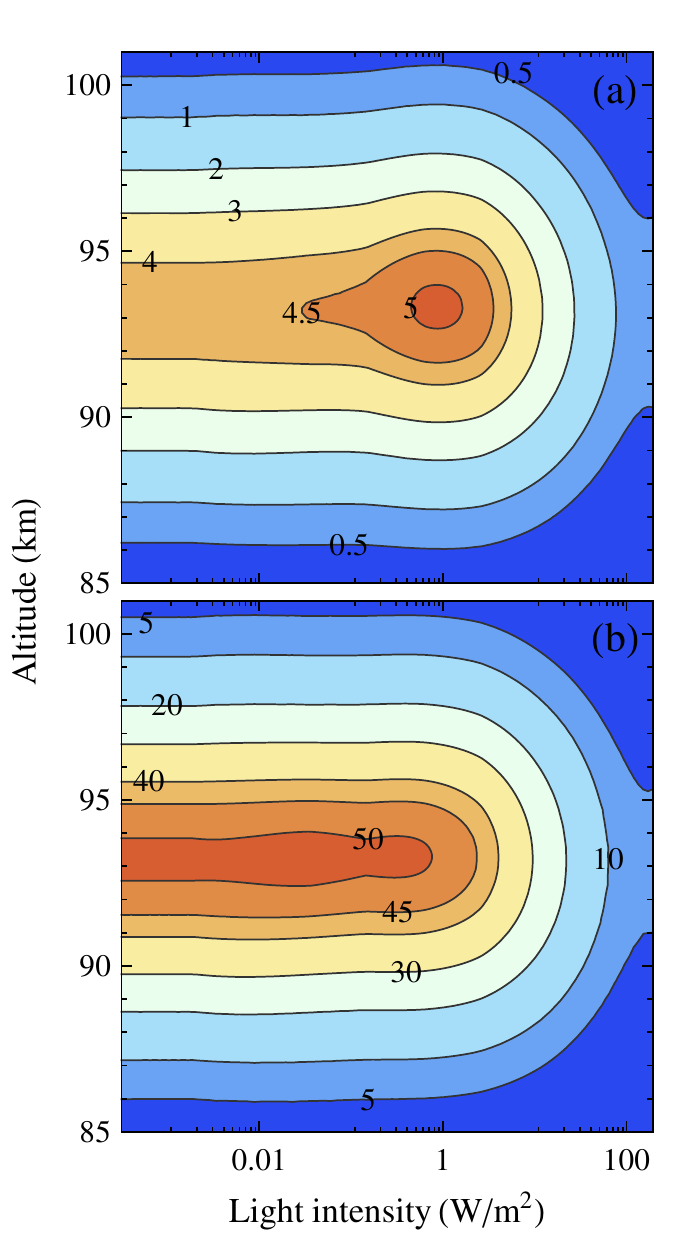}
\caption{(Color online) (a) Return flux $d\Phi(I,z)/d\ell/I$ in units of
$10^{9}$ photons/s/sr/m/(W/m$^2$) as a function of light
intensity and altitude. (b) Total flux
$d\Phi_\text{tot}(I,z)/d\ell/I$ in units of $10^{9}$
photons/s/m/(W/m$^2$), used to calculate optical absorption in
the sodium layer.
RochesterFig11.eps} \label{fig:FluxTablePlot}
\end{figure}

\begin{figure}[tbp]\centering
\includegraphics{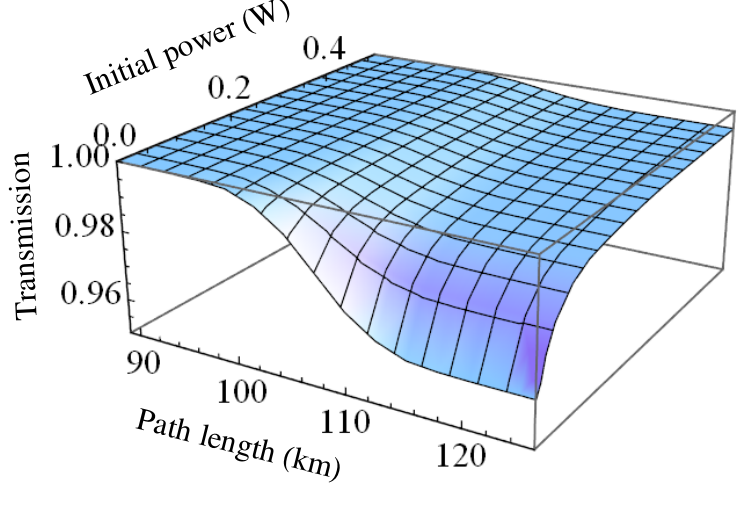}
\caption{(Color online) Transmission through the sodium layer as a function of
path length and initial light power in a pixel-sized beam.
RochesterFig12.eps}
\label{fig:UpwardTransmission}
\end{figure}

Here we illustrate the steps taken to perform the integration
of the photon return over the spatial extent of the beam.
First, we run the model a number of times assuming a 0.12"
pixel-sized spot to construct a table of the return flux
$d\Phi(I,z)/d\ell$ as a function of altitude and intensity
(Fig.\ \ref{fig:FluxTablePlot}a). At the same time we also
construct a table of the total fluorescence (Fig.\
\ref{fig:FluxTablePlot}b), which is used to calculate the
resonant absorption of light in the sodium layer (Fig.\
\ref{fig:UpwardTransmission}).

\begin{figure}[tbp]
\centering
\includegraphics{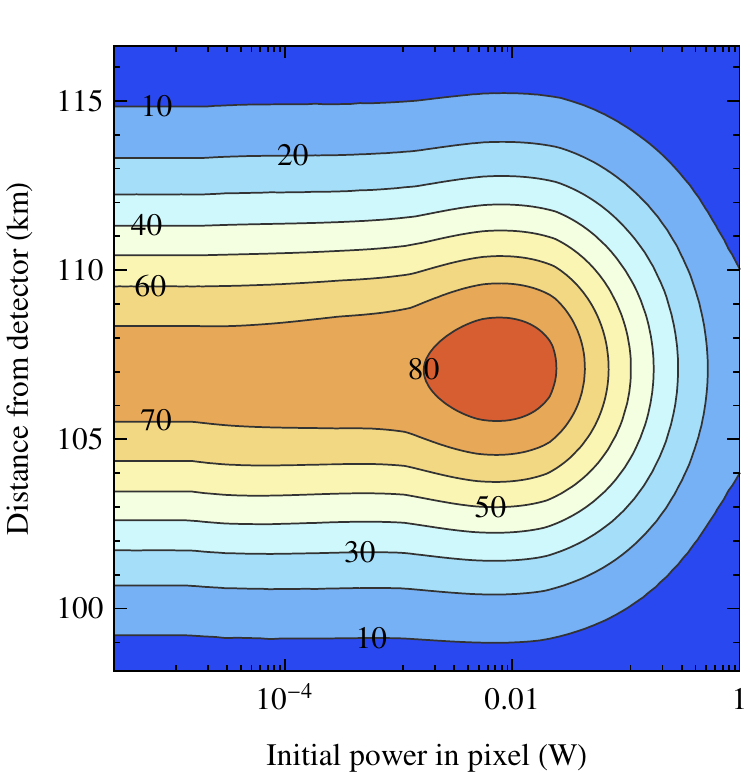}
\caption{(Color online) Normalized observed return flux per $d\ell$ from a
pixel-sized sub-beam, $d\Phi_\text{obs}(P_0,\ell)/d\ell/P_0$,
in units of photons/s/m$^2$/m/W.
RochesterFig13.eps} \label{fig:ObsFluxPlot}
\end{figure}

\begin{figure}[tbp]\centering
\includegraphics{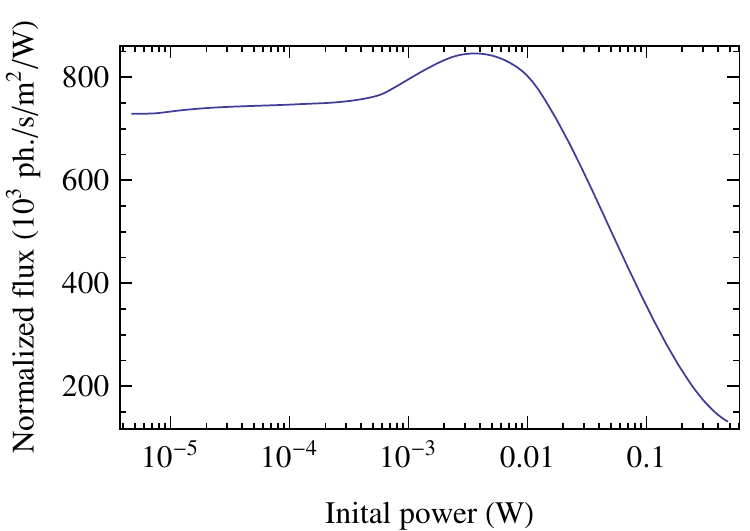}
\caption{(Color online) Total observed normalized return flux
$\Phi_\text{obs}(P_0)/P_0$ from a pixel-sized beam as a
function of initial launched light power in that pixel.
RochesterFig14.eps} \label{fig:IntObsFluxPlot}
\end{figure}

Accounting for resonant absorption on the uplink and downlink,
we can then construct a table of observed fluorescence at the
detector per unit path length in the sodium layer,
$d\Phi_\text{obs}(P_0,\ell)/d\ell$, as a function of distance
and initial power in each pixel (Fig.\ \ref{fig:ObsFluxPlot}).
Integrating over distance, we find the total return per pixel
as a function of initial power, $\Phi_\text{obs}(P_0)$ (Fig.\
\ref{fig:IntObsFluxPlot}). As seen in Fig.\
\ref{fig:IntObsFluxPlot}, the efficiency of return is constant
for low power (the linear regime), begins to rise as power is
increased, due to the beneficial effects of optical pumping,
and then falls off due to saturation effects, pumping to the
$F=1$ state, and recoil.

We then take the beam profile obtained from the geometrical
optics simulation with the appropriate adjusted Fried parameter
(Fig.\ \ref{fig:BeamAndFluxProfile}a), and apply
$\Phi_\text{obs}(P_0)$ to each pixel to obtain the return flux
spot profile (Fig.\ \ref{fig:BeamAndFluxProfile}b). Due to the
saturation effect, the central peak of the return flux profile
is compressed relative to that of the beam profile, increasing
the relative prominence of the dimmer features due to the
astigmatism seen at the top of the profile.

Summing over the pixels of the photon return flux profile gives
the final calculated value for the total return flux.

\subsection{Beam-integrated return flux}
\label{sec:integrationresults}

\subsubsection{Return flux as a function of zenith angle}

\begin{figure*}[tbp]
\centering
\includegraphics{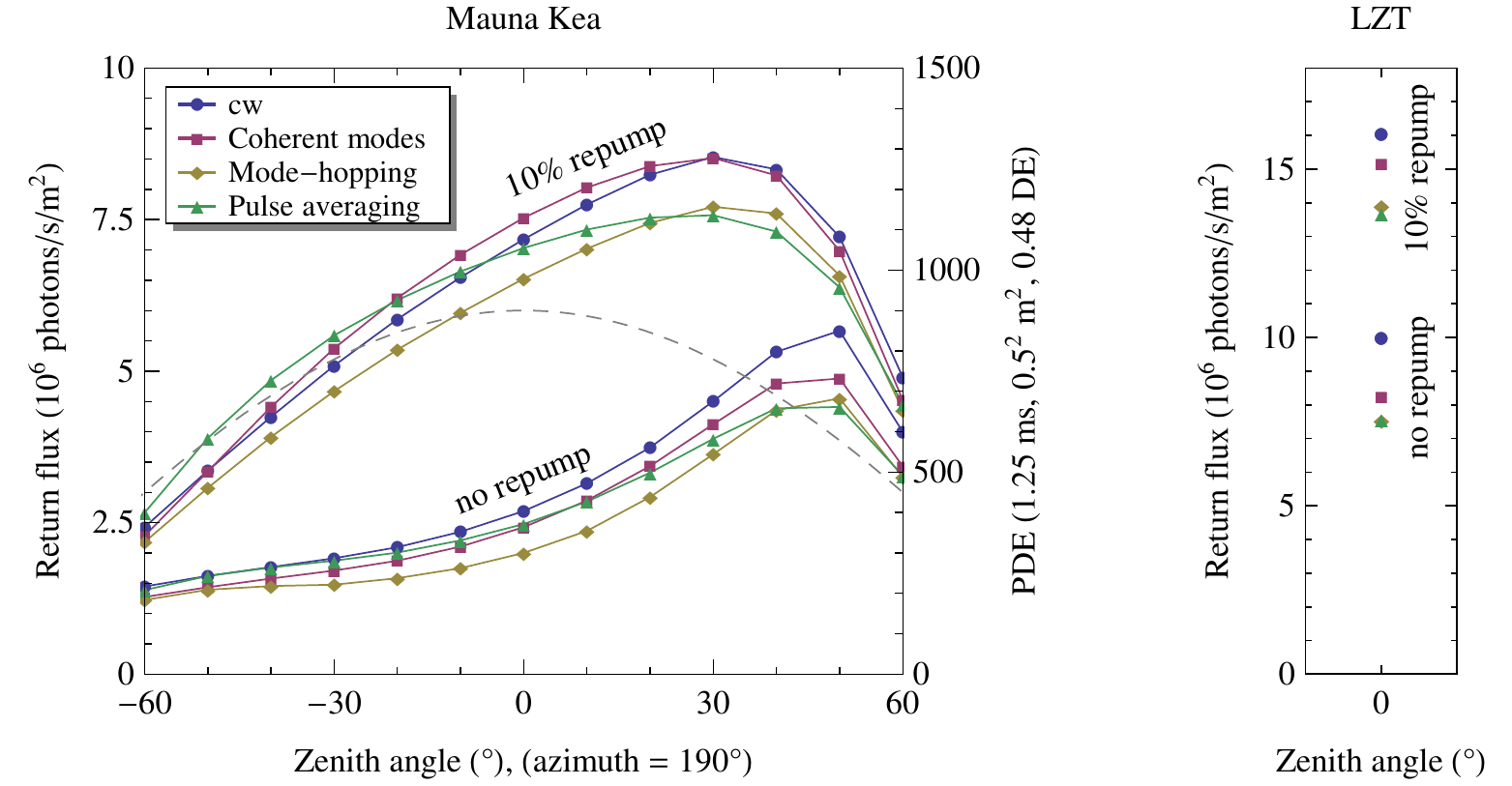}
\caption{(Color online) Beam-integrated return flux for Mauna
Kea (as a function of zenith angle) and LZT (at zenith), for
pulsed light using each of the three multimode methods and for
narrow-band cw light. Results are given for both no repump
light and for 10\% of the light power in the repump beam. The
vertical axis on the left side is in units of
$10^6\,\text{photons/s/m}^2$, while the axis on the right side
of the plot for Mauna Kea shows photo-detected electrons (PDE)
assuming a collection time of 1.25\,ms, collection area of
0.5\,m$^2$, and 48\% detection efficiency. The dashed gray line
shows the design requirement for 900 PDEs at zenith, combined
with the nominal dependance on the inverse of the airmass $X$
expected from purely geometrical considerations. (The optical
path length in the sodium layer increases with $X$, while the
detector collection solid angle falls as $X^{-2}$.) The higher
return at the LZT site is primarily due to the higher
throughput of the laser launch system and the different
direction of the local geomagnetic field compared to that at
the Mauna Kea site. RochesterFig15.eps}
\label{fig:IntegratedReturnVsAngles}
\end{figure*}

The total return flux integrated over the spatial extent of the
beam is shown in Fig.\ \ref{fig:IntegratedReturnVsAngles} as a
function of zenith angle for pulsed light, using each of the
three multimode methods, and also for narrow-band cw light.
Positive zenith angles correspond to $190^\circ$ azimuth, while
negative zenith angles correspond to $10^\circ$ azimuth.
Results are shown both for no repump light and for the standard
repump fraction (i.e., 10\% of the light power in the repump
beam, 80\% of the laser power in the D$_2$a light, and 10\%
lost).

The three methods for modeling the laser modes give results
that vary by $\sim$$15\%$; this may be taken as an indicator of
the level of uncertainty in the model due to incomplete
knowledge of the characteristics of the laser. (By running
calculations with finer grid spacings and stricter tolerances,
we have found that uncertainties due to numerical
approximations in the model are below 5\%.) At this level of
uncertainty, the narrow-band cw light provides approximately
the same photon return as the pulsed format. These results
indicate that both pulsed or cw light have the potential to
satisfy the TMT design requirements as long as repump light is
used, and will likely not satisfy the requirements if repump is
not used.

\subsubsection{Reversed-polarization repump}

\begin{figure}[tbp]
\centering
\includegraphics{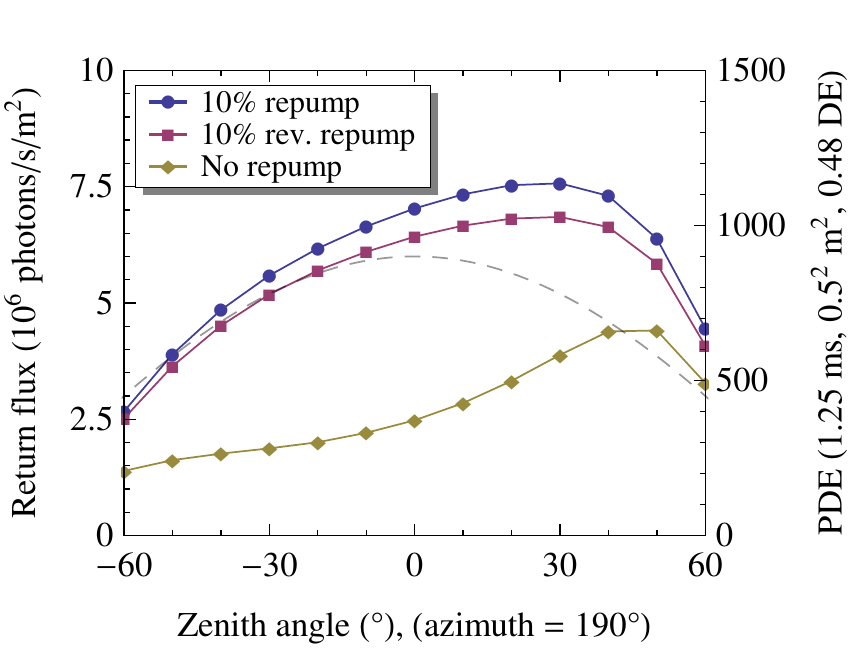}
\caption{(Color online) Beam-averaged return flux and PDE for a pulsed laser
without repump, with 10\% repump, and with 10\% repump with
opposite circular polarization as the D$_2$a
light.
RochesterFig16.eps}\label{fig:ReversedPolarizationRepump}
\end{figure}

For technical reasons, it may be easier to produce repump light
with opposite polarization to that of the main D$_{2}$a light.
Figure \ref{fig:ReversedPolarizationRepump} shows the effect of
reversing the polarization of the repump light. Results are
shown for a pulsed laser with no repump, with 10\% repump, and
with 10\% repump with opposite circular polarization to the
D$_2$a light, using the pulse-averaging multimode treatment.
Using opposite circular polarization for repumping incurs a
5--10\% penalty, depending on zenith angle, relative to
repumping with the same polarization. Because this is a
comparison of the same laser format and multi-mode model with
itself, an observed difference at this level may be numerically
significant.

\subsubsection{Dependence on spot size for LZT site}

To find the dependence of return flux on spot size for the LZT
site, the beam profile was scaled proportionally from the
nominal size 0.68" FWHM obtained from the geometrical optics
model using the standard Fried parameter $r_0=0.05$\,m. When
varying the beam size from 1/2 to 3 times the standard beam
size, the return flux varies from about 25\% below to 25\%
above the value obtained with the standard size.

\begin{figure}[tbp]
\centering
\includegraphics{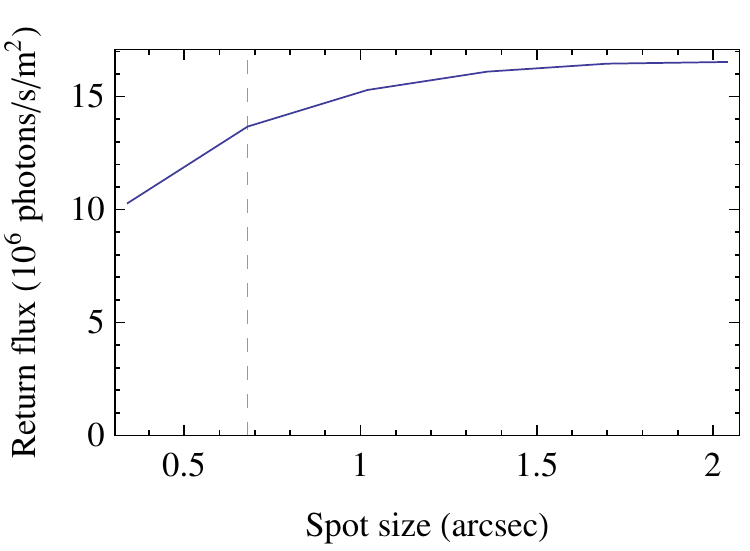}
\caption{(Color online) Return flux as a function of spot size (FWHM) at the
LZT site, using the pulse-averaging method. The dashed gray
line shows the spot size obtained from the geometrical optics
model using the standard Fried parameter $r_0=0.05$\,m. The
beam profile was directly scaled proportionally smaller and
larger to obtain the results for this plot.
RochesterFig17.eps}
\label{fig:LZTvsSpotSize}
\end{figure}

\subsection{Optimization}
\label{sec:optimization}

In this section, we plot the integrated return flux for the
Mauna Kea site as a function of zenith angle and various other
parameters for use in optimization. Results for pulsed and cw
lasers are given; for the pulsed format we use the
pulse-averaging multimode treatment.

\begin{figure}[tbp]\centering
\includegraphics{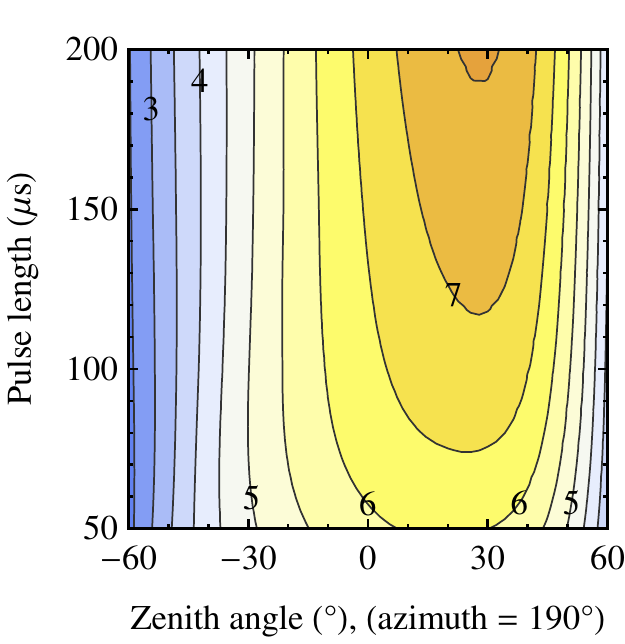}
\caption{(Color online) Return flux in units of $10^6$
photon/s/m$^2$ as a function of zenith angle and pulse length
for the Mauna Kea site. The repetition rate and average laser
power are held constant as the pulse length is varied.
RochesterFig18.eps} \label{fig:PulseAvgLengthDep}
\end{figure}

Figure \ref{fig:PulseAvgLengthDep} shows a contour plot of
return flux in units of $10^6$ photon/s/m$^2$ as a function of
zenith angle and pulse length, holding the repetition rate and
the average laser power constant. The standard amount of repump
light (of the same polarization as the D$_2$a light) is used.
The plot runs over the range of pulse lengths that may be
accessible to the TIPC laser. In this range, the plot shows
that longer pulse lengths are generally better.

\begin{figure}[tbp]\centering
\includegraphics{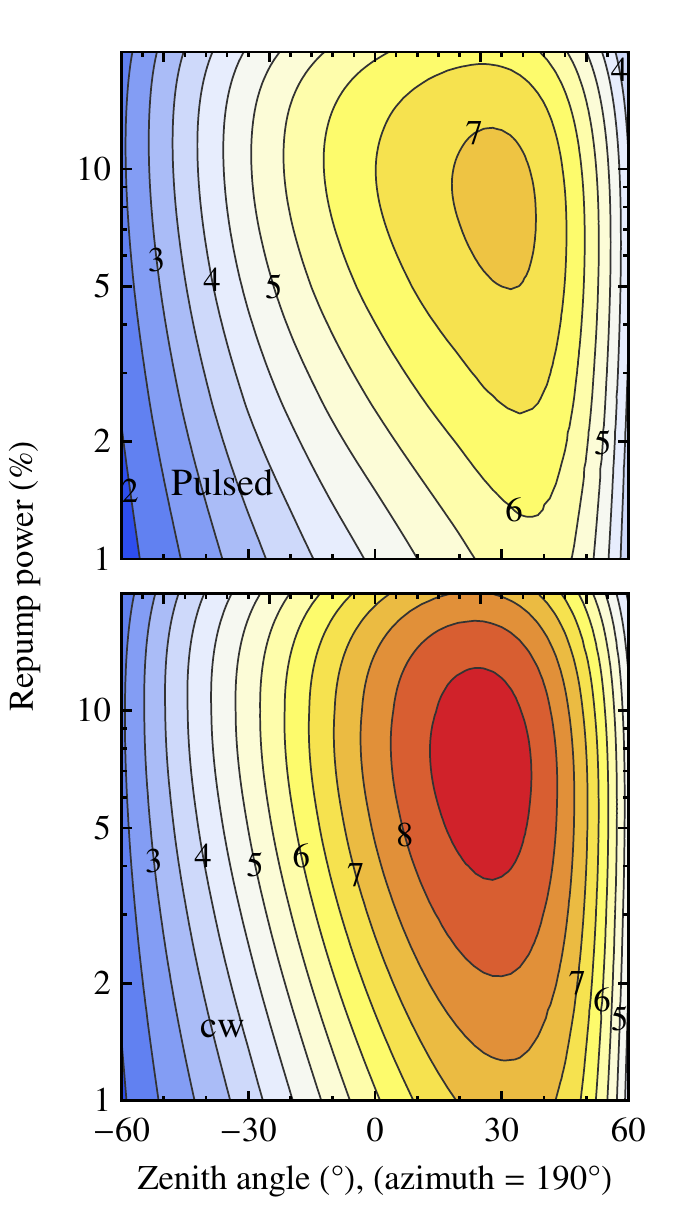}
\caption{(Color online) Return flux in units of $10^6$ photon/s/m$^2$ as a
function of zenith angle and repump power fraction for the
Mauna Kea site. Upper plot is for pulsed light, and
lower plot is for cw.
RochesterFig19.eps} \label{fig:ContourRepumpDep}
\end{figure}

Figure \ref{fig:ContourRepumpDep} shows a contour plot of
return flux as a function of zenith angle and repump power
percentage for the pulsed and cw cases. For Mauna Kea, it
appears that 10\% repump is optimal for both pulsed and cw
light.

\begin{figure}[tbp]\centering
\includegraphics{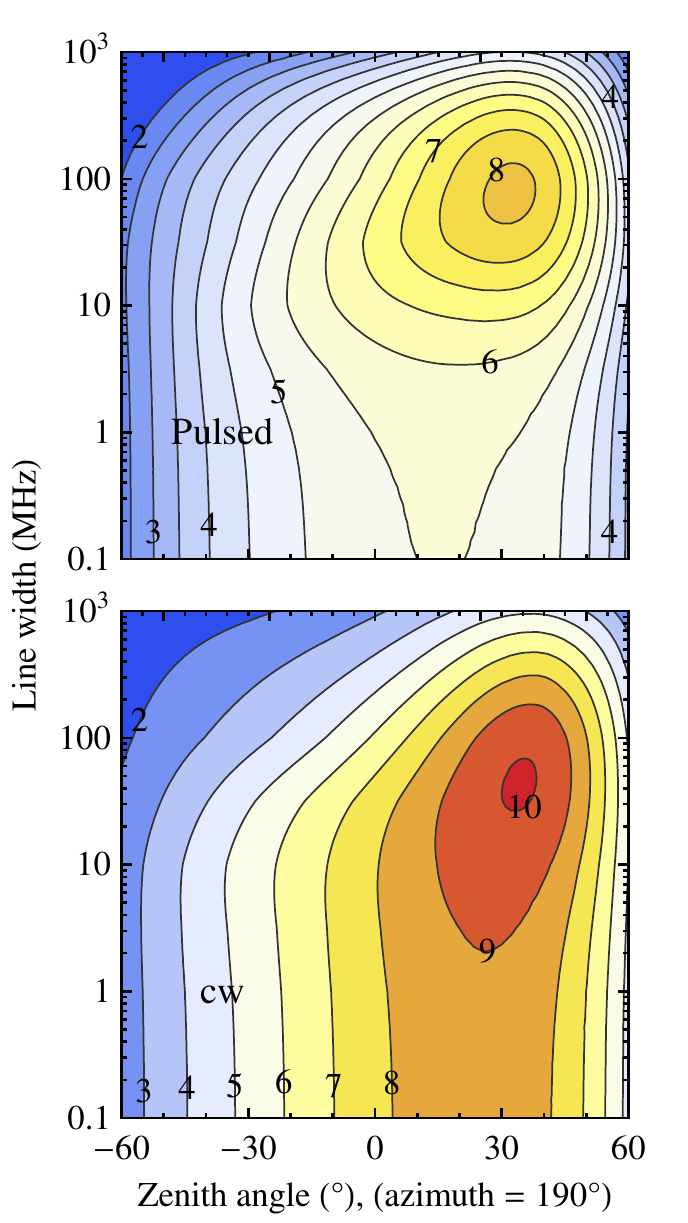}
\caption{(Color online) Return flux in units of $10^6$ photon/s/m$^2$ as a
function of zenith angle and line width of each mode for the
Mauna Kea site. Upper plot is for pulsed light, and lower plot
is for cw. The natural line width of the Na D$_2$ transition is
$\sim$10 MHz.
RochesterFig20.eps} \label{fig:ContourWidthDep}
\end{figure}

Figure \ref{fig:ContourWidthDep} shows a contour plot of return
flux as a function of zenith angle and the width of each laser
mode for the pulsed and cw cases. The standard amount of repump
light is used. The optimal width depends on the zenith angle,
or more precisely, the angle between the light propagation
direction and the magnetic field. When this angle is small, a
broader line width is advantageous, apparently because the
reduced effect of Larmor precession means that less light
intensity per velocity group is required to produce beneficial
atomic polarization. When the angle is larger, narrower laser
lines are optimal. For large angles, the optimum line width is
reasonably consistent with the value of $\sim$20\,MHz obtained
from the formula given in 
\cite{Hol2010} for the cw case. From these results we see that
increasing the line width of the modes of the pulsed laser
beyond the current assumed value of 15\,MHz would result in
reduced flux return for many observation directions.

\section{Conclusions}
\label{sec:conclusions}

Numerical simulations of photon return using both a pulsed
laser with specifications provided by TIPC and a narrow-band cw
laser have been performed for the Mauna Kea and LZT sites. The
results indicate that, if repump light is used, both the cw and
pulsed formats may be able to meet the design requirements for
photon return. Within the level of uncertainty in the model,
the cw laser format provides comparable return to the pulsed
format.

We have also investigated the dependence of the return flux on
parameters such as the pulse length, the fraction of laser
power dedicated to repumping, and the laser line width. For the
case of the laser line width, we found that the optimal value
has a strong dependence on the angle between the light
propagation direction and the magnetic field direction.

Future modeling work could use time-resolved measurements of
the mode structure of the TIPC laser to obtain more accurate
results. Another avenue of research could be to quantify the
impact of changing environmental conditions on the photon
return, for example, the sodium-layer density profile, optical
seeing conditions, and uncertainties in collision rates.

Finally, experimental validation of the model will greatly aid
future development. Published reports of comparison between
theoretical models and measurements for Na LGS systems have
been limited; there are some results for a cw LGS at the
Starfire Optical Range \cite{Hillman2008} and for a
transportable cw LGS developed by ESO \cite{Bonaccini2011}.
Significant information can be obtained even in the absence of
absolute sodium density measurements: by measuring the
dependence of the photon flux on experimentally variable
parameters, such as zenith angle, light power, light
polarization, and repump power fraction, the accuracy of
various aspects of the model may be verified, providing
information about the accuracy of the absolute photon return
values obtained from the simulation.

\section*{Acknowledgements}

The TMT Project gratefully acknowledges the support of the TMT
collaborating institutions.  They are the Association of
Canadian Universities for Research in Astronomy (ACURA), the
California Institute of Technology, the University of
California, the National Astronomical Observatory of Japan, the
National Astronomical Observatories of China and their
consortium partners, and the Department of Science and
Technology of India and their supported institutes. This work
was supported as well by the Gordon and Betty Moore Foundation,
the Canada Foundation for Innovation, the Ontario Ministry of
Research and Innovation, the National Research Council of
Canada, the Natural Sciences and Engineering Research Council
of Canada, the British Columbia Knowledge Development Fund, the
Association of Universities for Research in Astronomy (AURA)
and the U.S. National Science Foundation.

The authors also acknowledge the invaluable assistance of the
Technical Institute of Physics and Chemistry of the Chinese
Academy of Sciences.


\begin{thebibliography}{10}
\bibitem{curvaturenote} This formula for airmass dependence
    neglects the curvature of the earth, which
  is a reasonable assumption for the range of zenith angles ($\le60^\circ$)
  considered in this study.
\end{thebibliography}

\end{document}